# Direct Measurement of Electron Numbers Created at Near-Infrared Laser-Induced Ionization of Various Gases

A. Sharma[1], M. N. Slipchenko[1], K. A. Rahman[1], M. N. Shneider[2], and A. Shashurin[1]

[1] *Purdue University, West Lafayette, IN, USA*

[2] *Princeton University, Princeton, NJ, USA*

## Abstract

In this work, we present temporally resolved measurements of electron numbers created at photoionization of various gases by femtosecond laser pulse at 800 nm wavelength. The experiments were conducted in $O_2$, *Xe*, *Ar*, $N_2$, *Kr* and *CO* at room temperature and atmospheric pressure. Elastic microwave scattering was used to directly measure the electron numbers. Numbers of electrons in the range $3\cdot10^8$ to $3\cdot10^{12}$ electrons were produced by the laser pulse energies 100-700 μJ. After the laser pulse, plasma decayed on the time scale varied from 1 to 40 ns depending on the gas type and governed by two competing processes, namely, the creation of new electrons from ionization of the metastable atoms and loss of the electrons due to dissociative recombination and attachment to oxygen.

## Introduction

Broad research history of the laser-induced plasmas is related to studies of various nonlinear effects at laser beam propagation such as laser pulse filamentation, laser beam collapse, self-trapping, dispersion, modulation instability, pulse splitting etc.[1,2,3,4,5] These effects are various manifestations of the combined action of focusing Kerr nonlinearity (optical Kerr effect) and defocusing nonlinearity due to plasmas. Nowadays laser-induced plasmas find very wide application for plasma-assisted combustion, combustion diagnostics, laser-induced breakdown spectroscopy etc. [5]

Conventional techniques for diagnostics of laser-induced plasmas pose detrimental limitations. Specifically, the sensitivity of laser interferometry is limited to $n_e \geq 10^{16}$-$10^{17}$ cm$^{-3}$ due to the minimal measurable shifts of the interference fringes.[6,7,8] Numbers of semi-empirical methods for relative measurements of plasma density were proposed as well; however, all of them require absolute calibration based upon theoretically predicted values of plasma number density. Time-of-flight (TOF) mass spectrometer measurements of ion currents generated by laser-induced plasma have been conducted to measure photoionization rates.[1,9,10] The measurement relied on theoretical estimation of total number of electrons in the focal zone to conduct an absolute calibration of the system. Very recently, scattering of THz radiation from the laser-induced plasmas was proposed for spatially unresolved relative measurements of $n_e$ [10,11]. Other measurement techniques were proposed recently based on measurements of capacitive response times of system including a capacitor coupled with laser-induced plasma loaded inside.[11,12,13] These attempts to measure $n_e$ in laser-induced plasmas are characterized by various degrees of success and reliability of obtained data, but none of them provide the ultimate solution for absolute plasma density measurements until today.

Microwave scattering enables measurement of the total number of electrons ($N_e$) in the laser-induced plasma.[14,15,16] Very recently, electron numbers generated at femtosecond-laser ionization of air at atmospheric pressure was measured.[15] The idea of the approach is scattering the microwave radiation off the plasma volume in the quasi-Rayleigh regime (denoted as RMS in the following description), when the prolonged plasma volume is oriented along the linearly polarized microwaves and the plasma diameter is small compared to the spatial scale of the microwave field so that the incident microwave electric field is distributed uniformly across the entire plasma volume. "Quasi" is used to identify that scattering is equivalent to classical Rayleigh scattering if incident radiation is linearly polarized along the direction of the plasma object. In this case, the plasma electrons experience coherent oscillations in the incident microwave field and radiate a Hertzian dipole radiation pattern in a far field. Measurement of the scattered signal amplitude allows determining the total number of electrons in the plasma volume and average plasma density after appropriate calibration of the system with dielectric scatterers.

Thus, RMS technique demonstrated strong potential for absolute diagnostics of laser-induced plasmas. However, direct measurements of $N_e$ for a variety of other important atomic and molecular gases were not conducted before. This work is intended to fill this gap by conducting measurements of the number of electrons generated by femtosecond laser pulse in $O_2$, *Xe*, *Ar*, $N_2$, *Kr* and *CO*.

## Experimental Details

Schematic of the experimental setup is shown in Fig. 1. A regeneratively amplified Ti:Sapphire laser system (Solstice Ace; Spectra Physics, Inc.) was used as the fs-laser source having 126 fs Gaussian temporal width at 800 nm wavelength, which was focused into a positive pressure chamber using a plano-convex spherical lens of 1 m focal length. The laser was operated at 100 Hz repetition rate. The diameter of the beam incident on the lens was 7 mm. A half wave plate and thin film polarizer were used to control the irradiance of the beam, which was measured using a laser power meter (Gentec-EO XLP12-3S-H2-DO) placed after the lens. The positive pressure chamber was made in-house. Thin polyethene film was used for the walls of the chamber. It had small apertures for entry and exit of the laser beam and in order to maintain a continuous flow of the gas. The duration of the pulse was measured using single shot autocorrelation (TiPA, Light Conversion, Inc.)) and was logged throughout the experiment to ensure no drift in laser characteristics was taking place. For this work the measured beam waist and Rayleigh length were $w_o$=93.6 μm, $z_R$=26.9 mm, respectively.[15,16]

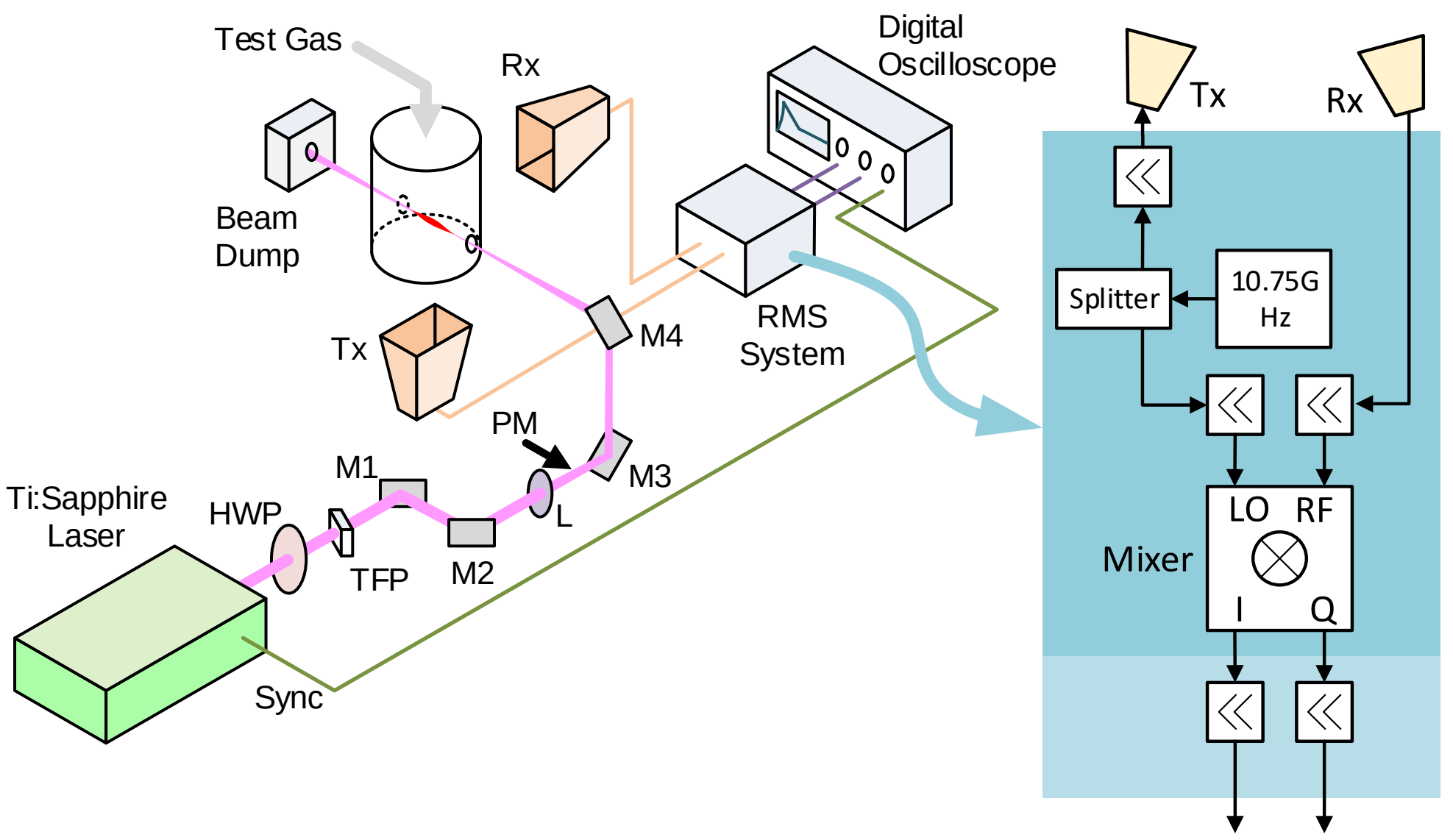

**Fig. 1. Femtosecond laser experimental setup and RMS homodyne measurement system.** Laser operates at 800 nm wavelength with 100 Hz repetition rate. The beam passes through a half wave plate (HWP) and thin film polarizer (TFP) which are used to control laser power measured using power meter (PM). The beam is focused using 1 m plano-convex (L) lens. The RMS system is calibrated using dielectric Teflon scatterers of know dimensions. For the dielectric scatterers, the output signal is recorded right after the mixer without additional amplification (32 dB power). Tested gases included: $O_2$, $Xe$, $Ar$, $N_2$, $Kr$, and $CO$.

RMS technique was used to measure the total number of electrons in the laser-induced plasma. Microwave radiation was used to irradiate the plasma. The polarization of the microwave electric field was kept along the direction of the elongated plasma volume. The plasma volume was located far enough from the microwave horn so that the incident electric field was uniform across the width of the plasma.[15,16] The plasma volume was polarized under the influence of the microwave field and radiated as a dipole antenna. Since the amplitude of radiation was uniform everywhere inside the plasma channel, the scattered radiation in the far field is analogues to that of the Hertzian dipole. This process is equivalent to elastic scattering of light in the Rayleigh regime when the wavelength of light is much greater than the size of the scatterer. The scattered signal from the plasma is proportional to the number of electrons inside the scattering volume. The RMS system was calibrated using a dielectric of known properties as scattering medium which enables absolute determination of electron numbers. Eq. (1) is the governing equation used for measuring the total number of electrons in the plasma. Where, $N_e$ - total number of electrons, $e$ - electron charge, $m_e$ - electron mass, $\nu$ – electron-gas collision frequency[18], $V$ - volume of the dielectric scatterer, $\omega$ - angular frequency of microwaves, $A$ – RMS system calibration coefficient, $\varepsilon_0$ - dielectric permittivity of vacuum, $\varepsilon$ - dielectric constant of the scatterer material and $U_{out}$ - output signal measured by the RMS system.

$$U_{out} = \begin{cases} A \cdot \dfrac{e^2}{m_e \nu} \cdot N_e - \text{ for plasma} \\ A \cdot V \cdot \varepsilon_0 (\varepsilon - 1)\omega - \text{for dielctric scatterer} \end{cases} \qquad (1)$$

For this experiment, a homodyne detection-based RMS system was used. Microwave source having frequency 10.75 GHz was used to irradiate the plasma volume. The output of the source was divided into two branches using a splitter. One branch carried the microwaves to the radiating horn after amplification. The other branch was connected to the LO port of the I/Q mixer. The scattered signal was received by a second horn, amplified and carried to the RF port of the mixer. The output of the mixer was again amplified and connected to a digital oscilloscope(Teledyne Lecroy HDO9304) to record the data.

The calibration of the RMS system was done by using Teflon dielectric bullets having a cylindrical shape with length 10 mm, diameter 3.175 mm and relative permittivity 2.10. The bullet was propelled using a pneumatic gun through the microwave field. The path of dielectric scatterer was chosen to be same as the path of the laser beam in the later experiments. Fig. 2 shows the output signal of the RMS system produced by Teflon bullet. The system calibration coefficient was found to be $A$ =6.72·10$^5$ $\mathrm{V\,\Omega\,m^{-2}}$.

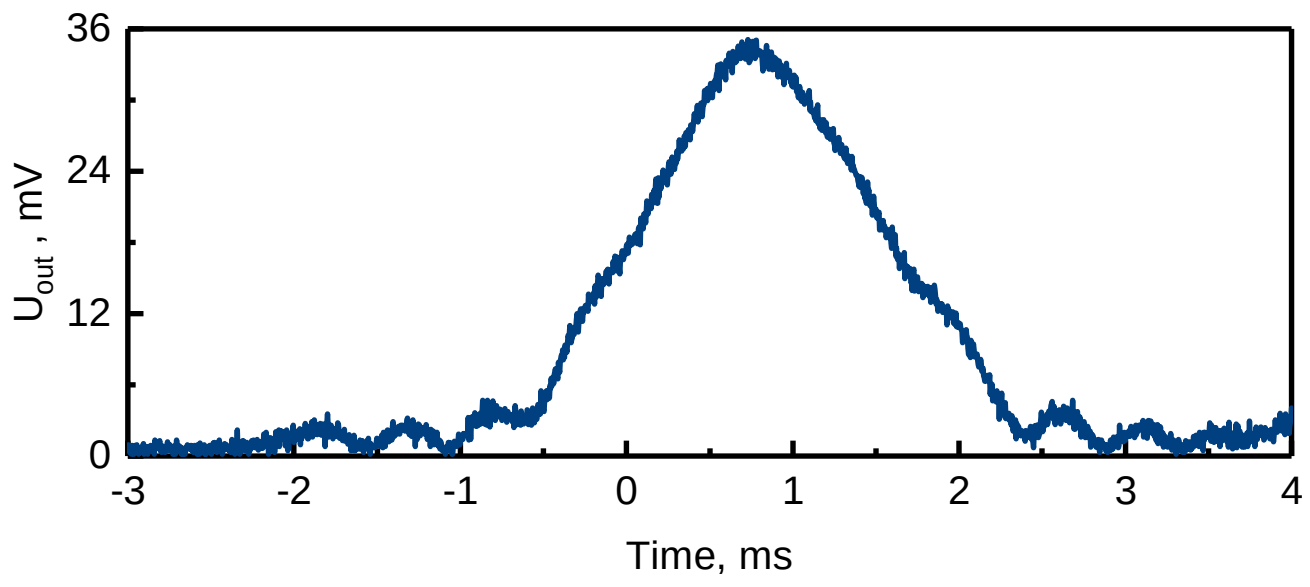

**Fig. 2. Response of the RMS system obtained from dielectric scatterer.** Temporal evolution of microwave signal scattered due to Teflon bullet was used for calibration of the RMS system.

Experimental setup shown in Fig. 3 was used to study the contribution of the non-linear processes in the laser beam propagation.[5] The system used the same 1000 mm lens as used in the setup shown in Fig. 1. The diverging beam after the focus was reflected from a pair of beam sampler onto a beam profiler. Each reflection reduced the intensity of the beam by 95%. $D4\sigma$ Beam diameter was measured by Newport LBP2-VIS2 Laser Beam Profiler. Measurements of the beam diameter were made for different pulse energies at two locations after the focus. The details of the measurements and analysis of the results are presented below.

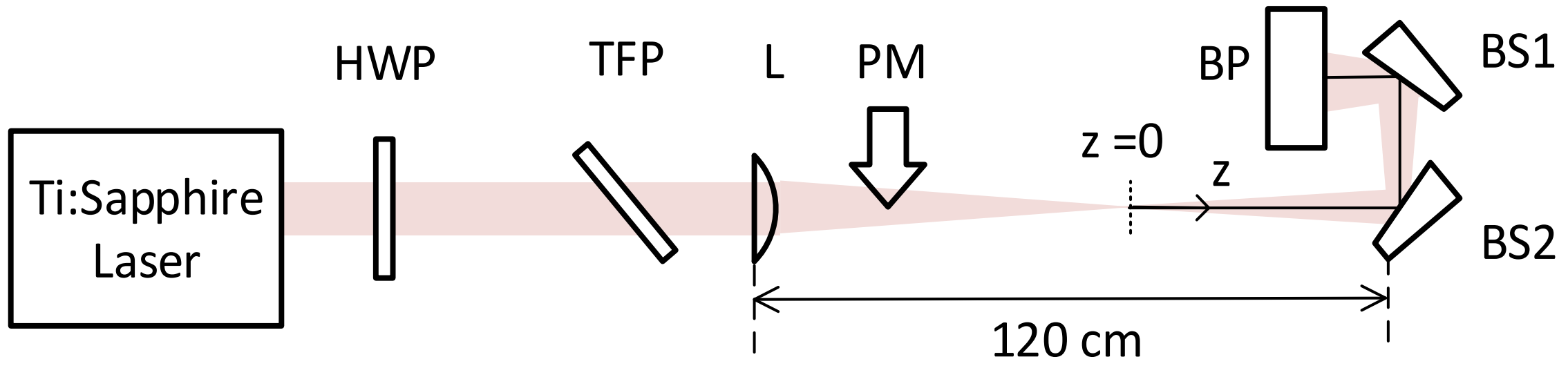

**Fig. 3. Experimental system used to study contribution of the non-linear processes in the laser beam propagation.** Beam profile was measured using laser beam profiler (BP) located at distance z=67 cm for different beam irradiances. A pair of wedge beam samplers (BS) was used to reduce the irradiance of the beam after focus to prevent saturation of BP.

## Results and Discussion

The moment of inception of non-linear effects was determined using the setup shown in Fig. 3. To this end, the beam radius $w$ was measured in ambient air at a location after the focus, namely $z$=67 cm for different laser pulse energies. Fig. 4 shows the dependence of beam radius vs. laser pulse energy and demonstrates three characteristic images of the beam profiles taken for $E_0$=80, 280 and 640 μJ. One can see that beam radius has saturated at 2.1mm with 5 % accuracy for $E_0 < 280\,\mu$J. However, there is rapid decrease in size for $E_0 > 280$μJ. The reduction in the beam size is attributed to non-linear Kerr effect near the focal region (Kerr effect dominates over the plasma non-linearity at low end of laser irradiances).[1,5] Thus, it can be concluded that threshold of non-linear effect inception in air corresponds to the laser pulse energy of $E_0$=280 μJ and corresponding irradiance, $I_0$=6.47·10$^{12}$ W/cm$^2$.

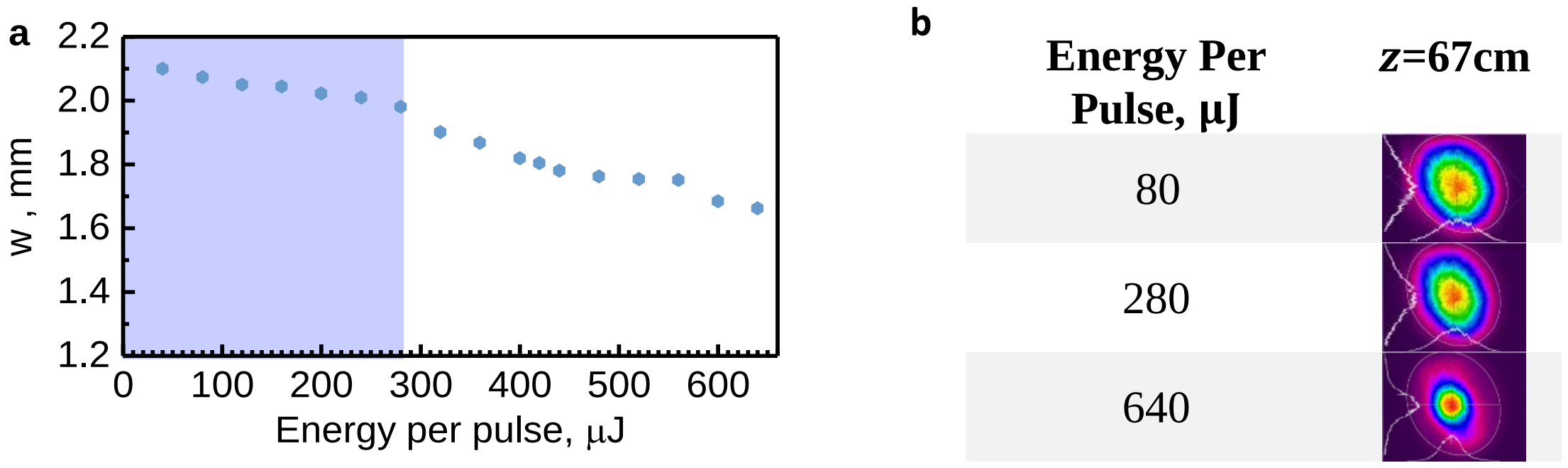

**Fig. 4. Measurements of inception of non-linear optical effects in air. (a)** Dependence of the laser beam radius $w = \sqrt{\frac{D4\sigma_x \cdot D4\sigma_y}{4}}$ at two locations after the focus ($z$=67 cm) vs. laser pulse energy $E_0$. **(b)** Images of the laser beam profiles for $E_0$=80, 280 and 640 μJ.

Based on the above findings, low laser pulse energies in the range of 220-300 μJ were used for determination of photoionization rate and electron number density in $O_2$ as detailed below. The non-linear effects are negligible in that range of energies and, therefore, the Gaussian beam distribution in vicinity of the focal plane can be used $I(r,z,t) = I_o\left(\frac{w_o}{w(z)}\right)^2 exp\left(-\frac{2r^2}{w(z)^2}\right) exp\left(\left(\frac{t-t^*}{\tau}\right)^2\right), w(z) = w_0\sqrt{1+\left(\frac{z}{z_R}\right)^2}$, where $\tau$ - characteristic temporal width of the beam ($\tau = \sqrt{2}\tau_{\text{Gauss}}$), $w_0$ – radius of beam waist and $z_R$ – Rayleigh length [15,16]. Cross-section of eight-photon ionization of oxygen molecule $\sigma_8$ was determined directly from the measurements of spatial and temporal characteristics of the laser beam ($\tau$, $w_o$ and $z_R$), measurements of $N_e$ by RMS system for different laser intensities, and using expression $N_e = \frac{231\pi}{1024\cdot 16}\sqrt{\frac{\pi}{8}}\sigma_8 n_0 \tau\pi w_0^2 z_R \cdot I_0{}^8$ following the methodology described in details in Ref. [15]. It was observed that proportionality law $N_e \propto I_0{}^8$ was satisfied with high accuracy in experiments. Multiphoton ionization rate of $O_2$ at atmospheric pressure (defined as $\nu = \sigma_8 I^8$) and plasma density in the centre ($n_{e0} = \frac{N_e}{\frac{231\pi}{1024\cdot 16}\pi w_0^2 z_R}$) are plotted in Fig. 5.

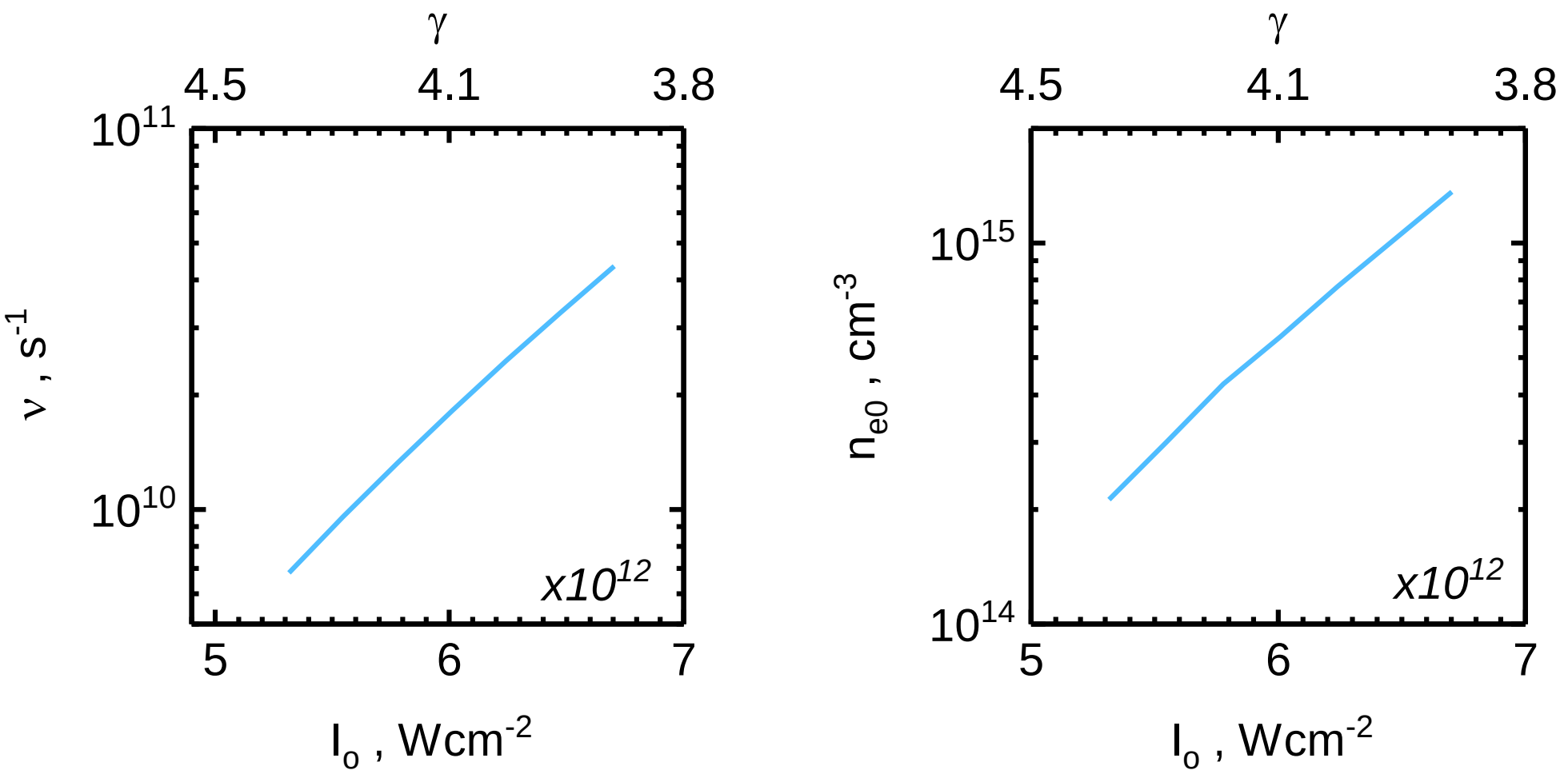

**Fig. 5 Multiphoton ionization of oxygen by femtosecond laser pulse (wavelength 800 nm) vs. irradiance at the waist$I_o$. The Keldysh parameter is shown on the top horizontal axis. (a)** Multiphoton ionization rate, **(b)** Plasma density vs. irradiance at the centre.

Fig. 6 presents log-log plot of the total number of electron in the plasma after laser pulse as a function of laser pulse energy in *Xe*, $O_2$, *Kr*, *CO*, $N_2$, and *Ar*. These experiments were conducted at atmospheric pressure and intensities when Gaussian beam intensity distribution does not apply due to non-linear phenomena. Therefore, in these experiments, only temporal dynamics of total number of electrons $N_e(t)$ was determined without making any additional assumptions on the intensity distribution in the laser beam. One can observe the following features. Firstly, the total number of electrons produced by the laser pulse in all gases increases as a power function of pulse energy. *Xe* is ionized the most to produce the highest number of electron from the same pulse energy compared to other gases in the experiment whereas $N_2$ produces the least. The energy requirement for electron generation increases going from left to right. Secondly, the ionization of *Xe* is ~2 orders of magnitude greater than $O_2$. Electron numbers generated in $N_2$ ($\varepsilon_i$ =15.6 eV) and *Ar* ($\varepsilon_i$ =15.7 eV) are almost the same for corresponding pulse energies (450-700 μJ) which is in agreement of their nearly identical ionization potentials. Ionization in *Kr* is 2 times greater than CO at lower pulse energies (~240 μJ) and the difference increases to 10 (~370 μJ). Thirdly, at higher pulse energies the electron production starts to saturate. The saturation sets in at ~180 μJ in *Xe*, 370 μJ in *Kr*, 320 μJ in $O_2$ and 400 μJ in *CO*. In $N_2$ and *Ar*, there is no onset of saturation for the pulse energies used in the experiments.

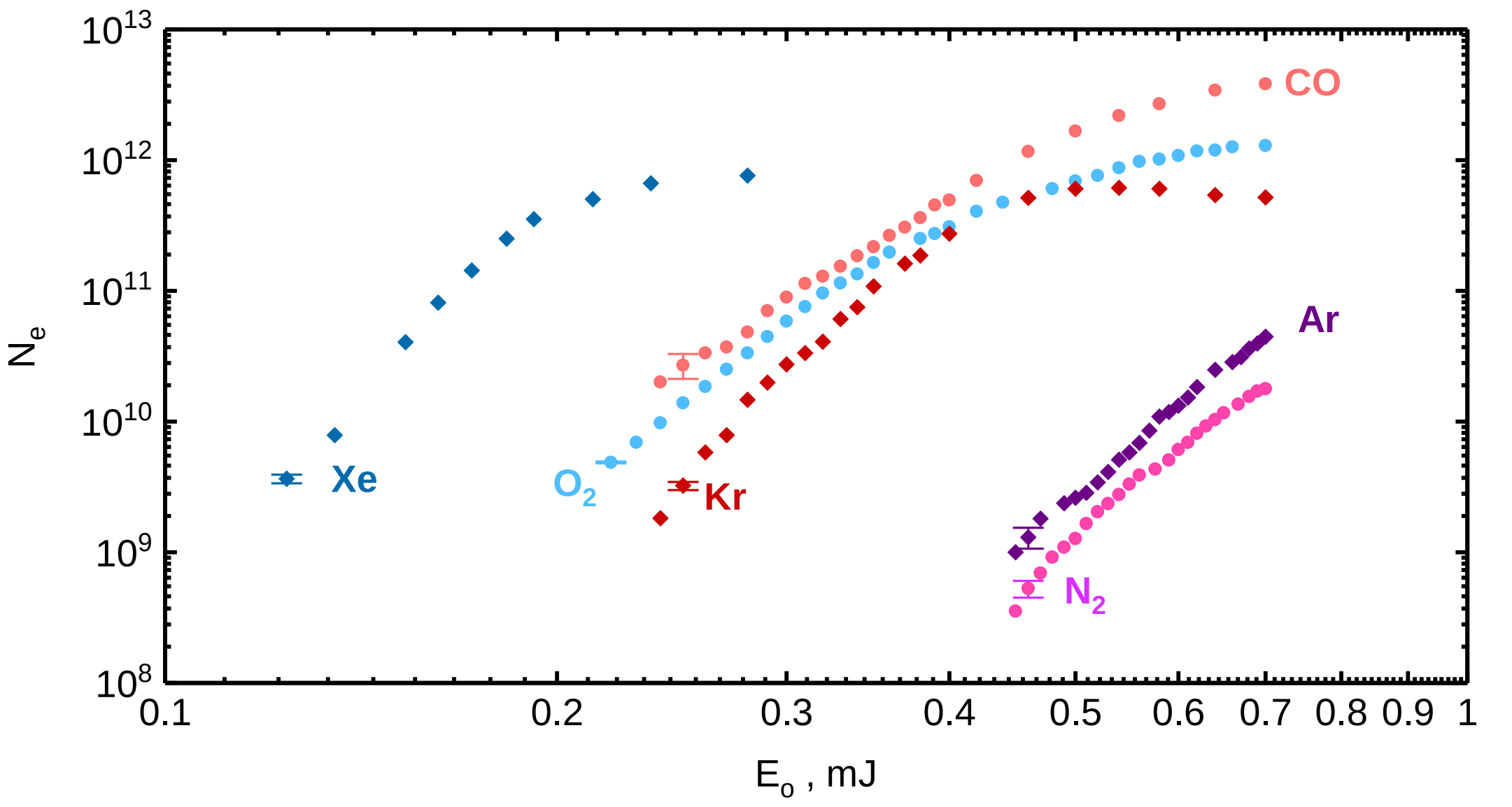

**Fig. 6. Total number of electrons generated by photoionization of different gases by laser pulse having Gaussian duration 126 fs and 7 mm diameter focused using 1,000 mm lens.**

Temporal decay of electrons is presented in Fig. 7. Electron decay in *Ar*, *Kr*, *Xe*, and $N_2$ can be explained by analysis of two competing processes, namely, creation of new electrons by ionization of the metastable atoms $M^*$ (where $M = Ar$, *Kr*, *Xe*, or $N_2$) at collision with electrons and loss of the electrons due to almost instant conversion of $M^+$ to $M_2^+$ and following dissociative recombination.

$$\frac{\partial n_e}{\partial t} = kn_e n_{M^*} - \beta n_e^2 \qquad (2)$$

In *Xe* and $N_2$ the recombination term remains dominant throughout the plasma decay $\beta n_e > kn_{M^*}$, so that number of plasma electrons always decreased with time. Anomalous temporal evolution of $N_e$ was observed in *Ar* and *Kr*; namely, one can see that $N_e$ continues to increase even after the laser pulse. This behavior can be potentially explained by high metastable density $n_{M^*}$ and thus dominant role of the ionization term immediately after the laser pulse $\beta n_e > kn_{M^*}$. Thus, $\frac{\partial n_e}{\partial t}$>0 immediately after the laser pulse and number of plasma electrons increased correspondingly. On later stage of the decay, number of metastable decreased and recombination term started to dominate causing further reduction of $N_e$ with time. Time corresponding to the peak value of $N_e$ can be used to estimate the plasma density in the laser-induced plasma. Indeed, $t_{peak} = 1/\beta n_e$, so using $\beta \approx 10^{-7}$ cm$^3$/s (Ref. 17) and experimentally observed $t_{peak}$~10 ns, the plasma density can be estimated as $n_e$ ~$10^{15}$ cm$^{-3}$.

$O_2$ and *CO* are characterized by high electron affinity, so additional electron loss mechanism, electron attachment, must be added to the right-hand side of the Eq. (2): $-\nu_a n_e$. In these gases, electron attachment mechanism dominates the overall electron loss and causes significantly faster $N_e$ decay observed on the time scale ~ 1 ns in the experiments shown in Fig. 7e and Fig. 7f. These decay times can be supported by the theoretical estimations. The rate of three-body attachment electron to oxygen molecule can be calculated using $\nu_a = k_{O^2} \cdot n_{O^2}^2$. The reaction constant $k_{O^2}$=$2.5{\cdot}10^{-30}$ cm$^6$/s (Ref. 17,19) gives the electron attachment rate $\nu_a \approx 10^9$ s$^{-1}$ and corresponding electron decay time $\tau_a = \frac{1}{\nu_a} \approx 1$ ns.

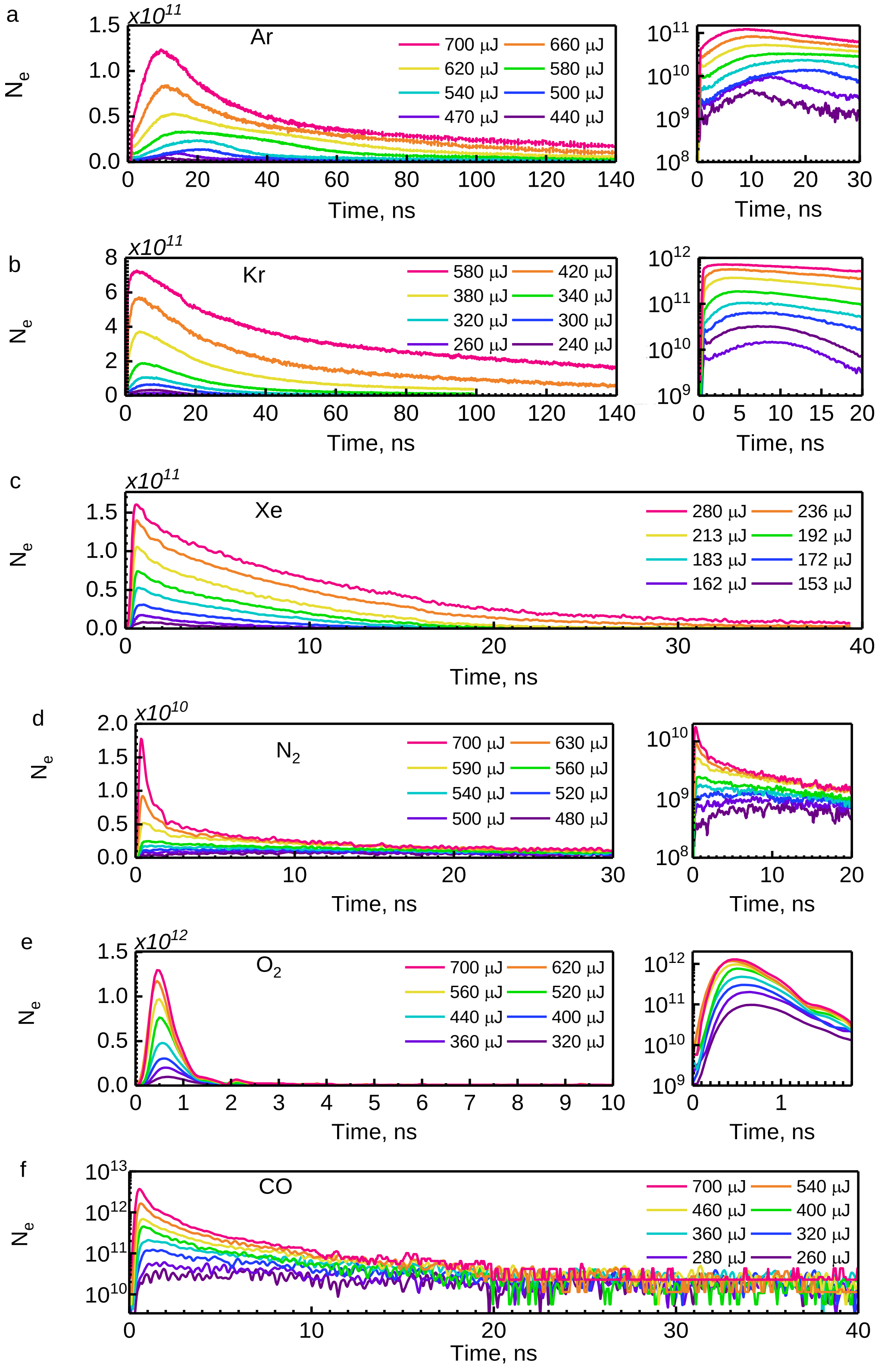

**Fig. 7. Temporal evolution of electron numbers in different gases (a)** *Ar*, **(b)** *Kr*, **(c)** *Xe*, **(d)** $N_2$ **(e)** $O_2$, **(f)** *CO*.

## Conclusion

In this work, we have successfully measured the total number of electrons created by focusing near-infrared femtosecond laser pulse for several gases at atmospheric pressure. The method was based on elastic scattering of microwaves off the plasma volume and absolute calibration of the microwave system using dielectric scatterers with known properties. Point of inception of non-linear optical effects in air was determined and photoionization rate in oxygen was determined for lower laser intensities ($< 6.47 \cdot 10^{12}$ W/cm$^2$). Electron decay after the laser pulse was consistent with competition of two processes, namely, creation of new electrons by ionization of the metastable atoms at collisions with electrons and electron loss due to dissociative recombination and attachment to oxygen.

## Acknowledgements

This work was partially supported by NSF/DOE Partnership in the Basic Plasma Science and Engineering program (Grant No. 1465061) and by U. S. Department of Energy (Grant No. DE-SC0018156).